# Effective multiband synthetic four-wave mixing by cascading quadratic processes

LI CHEN[1,2], ZHENG GE[1,2], SU-JIAN NIU[1,2], YIN-HAI LI[1,2], ZHAO-QI-ZHI HAN[1,2], YUE-WEI SONG[1,2], WU-ZHEN LI[1,2], REN-HUI CHEN[1,2], MING-YUAN GAO[1,2], MENG-YU XIE[1,2], ZHI-YUAN ZHOU[1,2,*], AND BAO-SEN SHI[1,2,*]

*1 CAS Key Laboratory of Quantum Information, University of Science and Technology of China, Hefei, Anhui 230026, China*
*2 CAS Center for Excellence in Quantum Information and Quantum Physics, University of Science and Technology of China, Hefei 230026, China*
*\* zyzhouphy@ustc.edu.cn*
*\* drshi@ustc.edu.cn*
*Authors to whom correspondence should be addressed: drshi@ustc.edu.cn*

**Abstract:** Four-wave mixing (FWM) is an important technique for supercontinuum and frequency comb generation in the mid-infrared band. Here we report simultaneous synthetic FWM in both the visible and mid-infrared bands by cascading quadratic nonlinear processes in a periodically poled lithium niobate (PPLN) crystal, which has a conversion efficiency that is 110 dB (at 3000 nm) higher than the FWM generated directly using third-order susceptibilities in bulk PPLN crystals. A general model of the proposed process is developed that shows full agreement with the experimental verification results. The frequency difference between the emerging frequency components can be tuned freely by varying the frequency difference between the dual pump lasers. Furthermore, by increasing the conversion bandwidth and the efficiency of the cascaded processes, it becomes feasible to generate frequency combs simultaneously in three bands, comprising the visible, near-infrared, and mid-infrared bands, via high-order cascaded processes. This work represents a route toward free-tuning multiband frequency comb generation with multi-octave frequency spanning that will have significant applications in fields including mid-infrared gas sensing, lidar, and high-precision spectroscopy.

With the ongoing development of laser sources, manipulation components, and detection devices for use in the mid-infrared band (2.5–25 μm), this wavelength band has attracted increasing attention in recent years [1-4]. The major focuses of the related research and potential applications in this band are described as follows: first, the vibration transition wavelengths of most molecules, including COx, NOx, and SOx, lie within this band and thus become fingerprints for molecular diagnosis and detection [5]; and second, the mid-infrared band contains two highly important atmospheric transmission windows (3–5 μm and 8–12 μm). Within these two transmission windows, the absorption losses of light caused by water vapor and carbon dioxide in the atmosphere are relatively low, and the atmospheric penetration capability in rainy and foggy weather is stronger than in the visible or near-infrared bands; these wavelengths can thus be used for free space optical communications [6] and lidar [7] applications. For applications including gas sensing, spectroscopy, and all-optical switching in the mid-infrared band [8-10], lasers with multiple frequency components are preferred. Four-wave mixing (FWM) is one of the most important physical mechanisms for use in generation of multifrequency components, including frequency combs [11-16] and supercontinua [17-22].

Generally, the efficiency of third-order nonlinear processes is very low because of the very weak third-order susceptibility $\chi^{(3)}$ of most materials, which limits the application of third-order nonlinear processes in bulk materials in many fields. Fortunately, the mechanism of cascading different nonlinear optical processes provides a promising method for the design of the nonlinear susceptibility properties of materials [23]. For example, we can use the lower-order $\chi^{(2)}$ process to construct an effective nonlinear process. By cascading quadratic nonlinearities, effective third-order nonlinear processes involving FWM, nonlinear phase shifts, third harmonics, and higher-order harmonics can be achieved [23-27]. The advantages of cascading processes include their high efficiency and extremely short response times, and cascading processes can thus be used in a wide range of application scenarios.

Cascaded quadratic processes have been demonstrated in bulk crystals and in micro/nano-optical structures using crystal materials that lack inversion symmetry, e.g., lithium niobate (LN) [28,29], $KTiOAsO_4$ [30], and aluminum nitride [31], or at the interfaces of centrally symmetrical media, including silicon dioxide [32] and silicon nitride [33]. Because of the excellent optical properties of LN, effective synthetic FWM processes have been studied widely on the basis of cascaded quadratic processes in bulk LN crystals (or waveguides) [34,35]. Generally, the cascaded schemes used for synthetic FWM can be divided into two groups. The first approach uses second harmonic generation and difference frequency generation (SHG+DFG) to synthesize third-order FWM effectively [34]. The second method involves synthesis of the FWM process using sum-frequency generation and DFG (SFG+DFG) [35]. When compared with the former approach, and because it is well known that SFG has a nonlinear coefficient that is twice that of SHG, the SFG+DFG process offers a higher conversion efficiency and the frequency tuning ability can be improved further by adjusting the wavelengths of the two pump lasers [36]. In previous works, the working bands of these two schemes were mainly concentrated within the visible and near-infrared bands and wavelength conversion was largely achieved in the telecom C band. Therefore, determination of a scheme that can generate FWM processes effectively in the mid-infrared bands is highly desirable.

In this paper, based on known synthetic FWM techniques [34,35], we develop a dual-pump cascading DFG and SFG (DPC-DFG+SFG) configuration. We generate FWM processes near the visible wavelength of 785 nm and the mid-infrared wavelength of 3000 nm in a PPLN crystal with high efficiency. A general coupling-wave equation model is constructed that explains all the experimental observations well. Using a combination of theoretical simulations and experiments, the conversion efficiency and the bandwidth of FWM in these visible and mid-infrared bands were explored. A frequency upconversion module was constructed to observe the spectrum in the mid-infrared band because of the lack of a suitable spectrum analyzer for use in that mid-infrared band. Finally, we discuss the possibility and the conditions required for generation of high-order FWM frequency sidebands using high-order cascaded processes.

In the theoretical part of the paper, we begin by explaining the principles of synthetic FWM based on cascading quadratic processes. As shown in Fig. 1(a)–(d), the synthetic FWM method based on the DPC-DFG+SFG cascaded quadratic processes in a PPLN crystal can be divided into four steps. The initial input light includes a continuous-wave (CW) visible signal laser beam S and two CW near-infrared pump laser beams P1 and P2 with slight wavelength differences. In the first step, mid-infrared laser beams DF1 and DF2 are generated through DFG using the signal light S and the pump lights P1 and P2 (these two processes are called DFG1 and DFG2, respectively). In the second step, the visible FWM signals SF1 and SF2 are created via SFG processes between the laser beams DF2 and DF1, and the pump beams P1 and P2 (these processes are called SFG1 and SFG2, respectively). In the third step, the mid-infrared FWM signals DF3 and DF4 are generated via DFG processes between the laser beams SF1 and SF2, and the pump beams P2 and P1 (these processes are called DFG3 and DFG4, respectively). In the fourth step, two groups of DFG processes contribute to the generation of near-infrared FWM signals: (i) the first group of DFG processes generates the near-infrared FWM signals DF5 and DF6 from DFG between the signal laser S and the signals DF3 and DF4, respectively (these processes are called DFG5 and DFG6, respectively); and (ii) the second group of DFG processes generate DF5 and DF6 via DFG processing between SF1 and SF2, and DF1 and DF2 (these processes are called DFG7 and DFG8, respectively). Because the higher-order cascaded processes of difference or sum-frequency processes have efficiencies that are two to three orders of magnitude lower than those of the four processes above, we can ignore the effects of the higher-order processes and consider the first-order FWM signals alone. However, as shown in the schematic diagrams, as long as the pump power and the crystal conversion bandwidth continue to increase, the higher-order sidebands are effectively generated in sequential cascaded quadratic processes, and this offers an effective method to extend the frequency components continuously in multiple bands. With appropriate phases locking the signal and pump lasers, frequency comb laser sources that span multiple octaves can be generated.

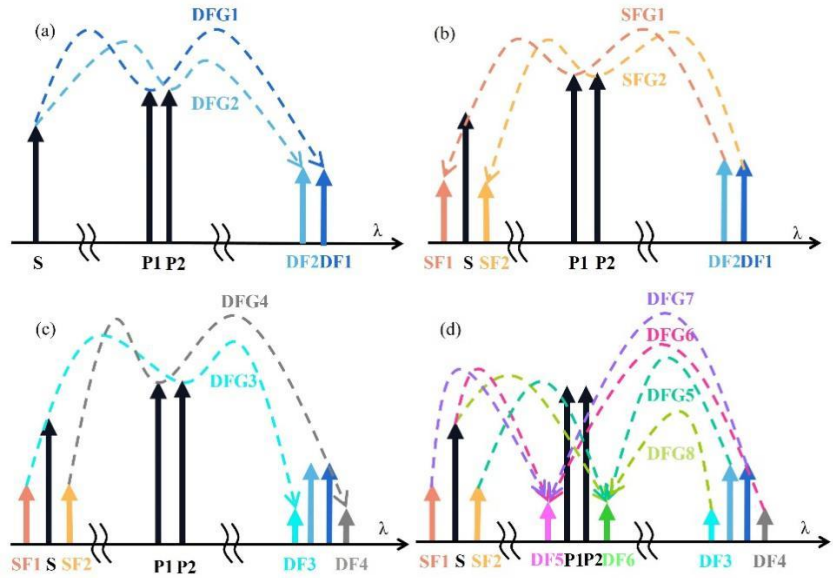

Fig. 1. Schematic diagram of the four steps (a)–(d) in the DPC-SFG+DFG process.

Second, we calculate the conversion efficiency of the cascaded quadratic process. Because the FWM in the near-infrared band (at approximately 1064 nm) involves three cascaded quadratic processes and lower efficiency than before, the higher-order cascaded quadratic process shown in Fig. 1(d) is not considered here.

The cascaded quadratic processes in the PPLN crystal can be described using the following coupling wave equations [37]:

$$\begin{cases}
\dfrac{dE_{p1}}{dz} = \dfrac{2id_{eff}\omega_{p1}}{n_{p1}c}\left[E_s E_{DF1}^* \exp(-i\Delta k_{DFG1}z) + E_{SF1}E_{DF2}^* \exp(-i\Delta k_{SFG1}z) + E_{SF2}E_{DF4}^* \exp(-i\Delta k_{DFG4}z)\right] \\[4pt]
\dfrac{dE_{p2}}{dz} = \dfrac{2id_{eff}\omega_{p2}}{n_{p2}c}\left[E_s E_{DF2}^* \exp(-i\Delta k_{DFG2}z) + E_{SF2}E_{DF1}^* \exp(-i\Delta k_{SFG2}z) + E_{SF1}E_{DF3}^* \exp(-i\Delta k_{DFG3}z)\right] \\[4pt]
\dfrac{dE_s}{dz} = -\dfrac{2id_{eff}\omega_s}{n_s c}\left[E_{DF1}E_{p1}\exp(i\Delta k_{DFG1}z) + E_{DF2}E_{p2}\exp(i\Delta k_{DFG2}z)\right] \\[4pt]
\dfrac{dE_{DF1}}{dz} = \dfrac{2id_{eff}\omega_{DF1}}{n_{DF1}c}\left[E_s E_{p1}^* \exp(-i\Delta k_{DFG1}z) + E_{SF2}E_{p2}^* \exp(-i\Delta k_{SFG2}z)\right] \\[4pt]
\dfrac{dE_{DF2}}{dz} = \dfrac{2id_{eff}\omega_{DF2}}{n_{DF2}c}\left[E_s E_{p2}^* \exp(-i\Delta k_{DFG2}z) + E_{SF1}E_{p1}^* \exp(-i\Delta k_{SFG1}z)\right] \\[4pt]
\dfrac{dE_{SF1}}{dz} = -\dfrac{2id_{eff}\omega_{SF1}}{n_{SF1}c}\left[E_{DF2}E_{p1}\exp(i\Delta k_{SFG1}z) + E_{DF3}E_{p2}\exp(i\Delta k_{DFG3}z)\right] \\[4pt]
\dfrac{dE_{SF2}}{dz} = -\dfrac{2id_{eff}\omega_{SF2}}{n_{SF2}c}\left[E_{DF1}E_{p2}\exp(i\Delta k_{SFG2}z) + E_{DF4}E_{p1}\exp(i\Delta k_{DFG4}z)\right] \\[4pt]
\dfrac{dE_{DF3}}{dz} = \dfrac{2id_{eff}\omega_{DF3}}{n_{DF4}c} E_{SF1}E_{p2}^* \exp(-i\Delta k_{DFG3}z) \\[4pt]
\dfrac{dE_{DF4}}{dz} = \dfrac{2id_{eff}\omega_{DF4}}{n_{DF4}c} E_{SF2}E_{p1}^* \exp(-i\Delta k_{DFG4}z)
\end{cases} \quad (1)$$

where $E_i$ ($i=p1, p2, S, SF1, SF2, DF1, DF2, DF3, DF4$) are the electrical amplitudes, $n_i$ are the refractive indices of the beams inside the crystal, $\omega_i$ represent the frequencies of the light beams, $d_{eff}$ is the effective nonlinear coefficient, $c$ is the velocity of light, and $\Delta k_i$ ($i=SFG1, SFG2, DFG1, DFG2, DFG3, DFG4$) are the phase mismatches in the cascade process. The expressions for the phase mismatches $\Delta k_i$ are given as follows:

$$\begin{aligned}
\Delta k_{DFG1} &= k_s - k_{DF1} - k_{p1} - 2\pi/\Lambda, & \Delta k_{DFG2} &= k_s - k_{DF2} - k_{p2} - 2\pi/\Lambda \\
\Delta k_{SFG1} &= k_{SF1} - k_{DF2} - k_{p1} - 2\pi/\Lambda, & \Delta k_{SFG2} &= k_{SF2} - k_{DF1} - k_{p2} - 2\pi/\Lambda \\
\Delta k_{DFG3} &= k_{SF1} - k_{DF3} - k_{p2} - 2\pi/\Lambda, & \Delta k_{DFG4} &= k_{SF2} - k_{DF4} - k_{p1} - 2\pi/\Lambda
\end{aligned} \quad (2)$$

$\Lambda$ is the poling period here, and $k_i = 2\pi n_i/\lambda_i$ are the wave vectors of the signal, pump, and cascaded light beams. By considering the slowly varying amplitude approximation and the undepleted pump approximation, we then define and derive the FWM conversion efficiency in the visible band as generated using the first-order cascaded process as follows:

$$\eta_{\text{visible}} = P_{SF2}/P_S = \dfrac{128\pi^4 d_{eff}^4 P_{p1} P_{p2} L^2}{\varepsilon_0^2 c^2 n_S n_{SF2} n_{DF1}^2 \lambda_{DF1}^4 \lambda_{p1}\lambda_{p2}} h_0(0,\xi_0) h_1(0,\xi_1) \quad (3)$$

Here, $\xi_0 = L/2Z_{p1}$ is the focusing parameter of the pump light beam and the expression for $h_0(0,\xi_0)$ can be written as:

$$h_0(0,\xi_0) = \frac{1}{\xi_0}\int_{-\xi_0}^{\xi_0}dx\int_{-\xi_0}^{\xi_0}dy\frac{e^{-i\sigma_0(x-y)}}{(1+ix)(1-iy)[2+\frac{i(x-y)}{\beta_0}]+\alpha_0(1+\frac{ix}{\beta_0})(1-\frac{iy}{\beta_0})[2+i(x-y)]} \quad (4)$$

where $\alpha_0 = \omega_{0s}^2/\omega_{op1}^2$, $\beta_0 = Z_{0s}/Z_{0p1} = n_s\lambda_{p1}\omega_{0s}^2/n_{p1}\lambda_s\omega_{0p1}^2 = n_s\lambda_{p1}\alpha/n_{p1}\lambda_s$, and the spatial phase-mismatching parameter $\sigma_0 = \Delta k_{DFG1}Z_{p1}$ and $h_1(0,\xi_1)$, $h_2(0,\xi_2)$, and $h_3(0,\xi_3)$ have similar definitions.

Next, we define and derive the FWM conversion efficiency in the mid-infrared band that is generated by the second-order cascaded process as follows:

$$\eta_{MIR} = P_{DF4}/P_{DF2} = \frac{64\pi^4 d_{eff}^4 P_{p1}^2 L^2 n_{DF2}\lambda_{DF2}^2 h_0(0,\xi_0)h_1(0,\xi_1)h_3(0,\xi_3)}{\varepsilon_0^2 c^2 n_{SF2}^2 n_{DF1}^2 n_{DF4}\lambda_{DF1}^4 \lambda_{DF4}^2 \lambda_{p1}^2 h_2(0,\xi_2)} \quad (5)$$

Here, $\xi_0 = \xi_3 = L/2Z_{p1}$ and $\xi_1 = \xi_2 = L/2Z_{p2}$ are the focusing parameters of the pump light beam, $L$ is the crystal length, $\lambda_i$ are the wavelengths of the light beams, and $\varepsilon_0$ is the permittivity of a vacuum. The reader should refer to the Supplementary Information for further details of the derivation of the above. With these analytical expressions for the FWM conversion efficiency in the visible and mid-infrared bands, we can then obtain the relationship between the FWM conversion efficiency and the pump power, along with the detuning of the two pump wavelengths.

For this experimental part, a schematic diagram of the experimental setup used for the DPC-DFG+SFG scheme in the PPLN crystal is shown in Fig. 2. The signal laser beam S for the cascaded quadratic process comes from a diode laser (TOPTICA pro, Graefelfing, Germany), the spatial mode of which was later optimized by passing the beam through a single-mode fiber. The two pump laser beams are provided by two diode lasers and are enhanced by fiber amplifiers; the wavelength of pump laser P1 is fixed at 1063.9 nm, and the wavelength of pump laser P2 can be tuned continuously near 1063.9 nm. After the two pump beams are combined using a beam splitter (BS), they overlap with the signal light beam after dichroic mirror DM1 (with high reflection at 785 nm and high transmission at 1064 nm); each interacting beam was set to have vertical polarization by wave plates located before the nonlinear crystal, and thus satisfied the type-0 phase-matching condition in the PPLN. In the PPLN crystal, the beam waists of the pump lasers and the signal laser are 65 μm and 50 μm, respectively, at the focal point. The temperature of each crystal was controlled using custom-made temperature controllers, which have temperature stability of ±2 mK. After beam interaction in the PPLN crystal, a 3000T/785R dichroic mirror (DM2) and a 3000T/1064R dichroic mirror (DM3) were used to filter out the laser beams in their different wavelength bands. The FWM spectrum shown in the visible band was observed directly using an optical spectral analyzer (OSA).

Because no mid-infrared OSA can be obtained in our laboratory, the FWM signal in the mid-infrared band was first converted into the visible band through frequency upconversion, and the converted signal was then detected using a visible OSA. We take advantage of the wide bandwidth conversion property of the chirp PPLN (CPPLN) crystal to construct the upconversion system. Before the CPPLN, a 1400 nm long-pass filter is used to remove the residual signal laser and pump laser beams from the output laser signals passing through the first crystal. The pump laser beam P2 is first split by the polarizing beam splitter (PBS), which overlaps with the signal light after DM4 of the 3000T/1064R dichroic mirror, and is then focused by a lens group with beam waists of 50 μm and 60 μm. The upconversion laser beam from the conversion module is acquired by removing the pump laser and signal laser beams using filters and is then connected to the visible OSA to analyze the spectrum of the mid-infrared FWM process.

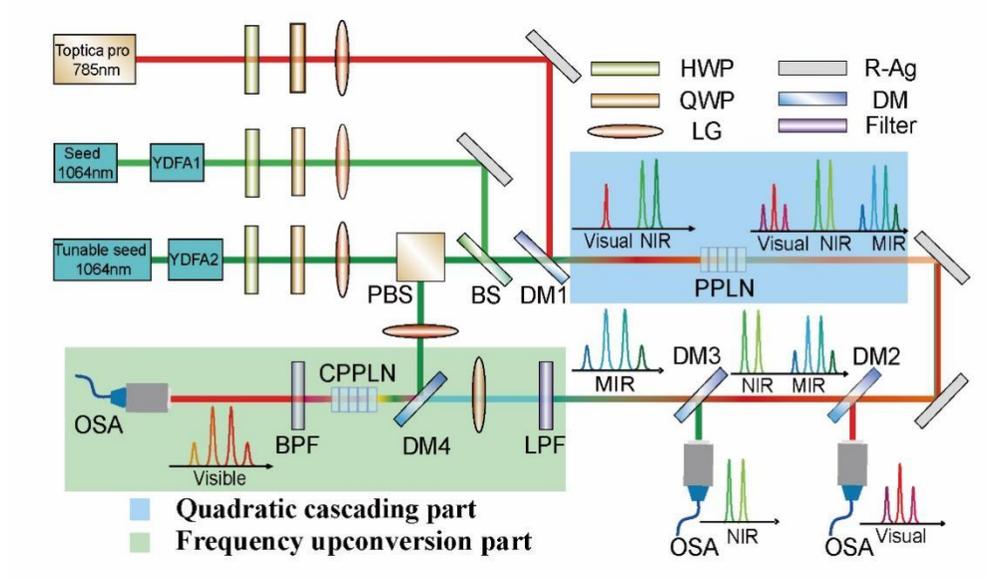

Fig. 2. Schematic illustration of the experimental setup used for the DPC-DFG+SFG system. PPLN: periodically poled lithium niobate crystal with dimensions of $40\ mm \times 10\ mm \times 0.5\ mm$ (height); CPPLN: chirp PPLN crystal that has dimensions of $40\ mm \times 3\ mm \times 2\ mm$ (height); YDFA: Yb3+-doped fiber amplifier; HWP: half-wave plate; QWP: quarter-wave plate; LG: lens group; R-Ag: Ag mirrors; BS: beam splitter operating at 1064 nm; PBS: polarizing beam splitter operating at 1064 nm; DM1: 785T/1064R dichroic mirror; DM2: 3000T/785R dichroic mirror; DM3: 3000T/1064R dichroic mirror; DM4: 3000T/1064R dichroic mirrors; LPF:1400 nm long-pass filter; BPF: 800–40 nm bandpass filter; OSA: optical spectral analyzer.

In the results section, to explore the cascaded quadratic process in the PPLN crystals, it is necessary to characterize the quantum efficiency, wavelength, and temperature tuning characteristics of the DFG process in the PPLN and the SFG process in the CPPLN when using the same 1064 nm pump light. We use the DFG process in the PPLN as an example to demonstrate the nonlinear characteristics of the PPLN crystal. Before it reached the PPLN crystal, the pump laser power was 8 W, and the signal laser power was 1 W. The DFG process generates a 30.5 mW mid-infrared laser beam at a wavelength of 3000 nm. As shown in Fig. 3(a), at the low conversion efficiency, the nonlinear process satisfies the undepleted pump approximation, and the output DFG laser power is directly proportional to the pump light power. By considering the 9.68% power loss caused by the subsequent filters and the dichroic mirrors, the power efficiency of the DFG process as determined using $\eta_{power} = P_{3000} / P_{785}$ was 3.38% for a pump power of 8 W, and the corresponding quantum conversion efficiency (QCE), as defined by $\eta_{quantum} = \eta_{power} \lambda_{3000} / \lambda_{785}$, was 12.92%. Furthermore, to test the wavelength bandwidth of the crystal, the temperature of the crystal was fixed at 39°C, we varied the signal wavelength to measure the normalized QCE (NQE), and the conversion bandwidth of the DFG process was 0.21 nm at 785.16 nm. Through further testing of the relationship between the best matching temperature and the signal wavelength, the wavelength-temperature adjustment coefficient of the PPLN crystal was found to be 0.067 nm/°C at 785.16 nm, and the measured crystal characteristic curves were as shown in Fig. 3(b) and 3(c). Similarly, to characterize the properties of the upconversion module, the SFG process in the CPPLN crystal was tested at a signal laser power of 2 mW, with results as shown in Fig. 3(d)–(f). The relevant parameters for the pump power of 8 W are given in Table I.

Table I. Comparison between characteristic parameters of DFG in PPLN and SFG in CPPLN

| Crystals | Power efficiency at 8 W pump power | QCE | Conversion bandwidth(nm) | Temperature coefficient(nm/°C) |
|---|---|---|---|---|
| PPLN | 3.38% | 12.92% | 0.21 @785.16 nm | 0.067 @785.16 nm |
| CPPLN | 4.95% | 1.29% | 8.8 @3000 nm | 0.97 @3000 nm |

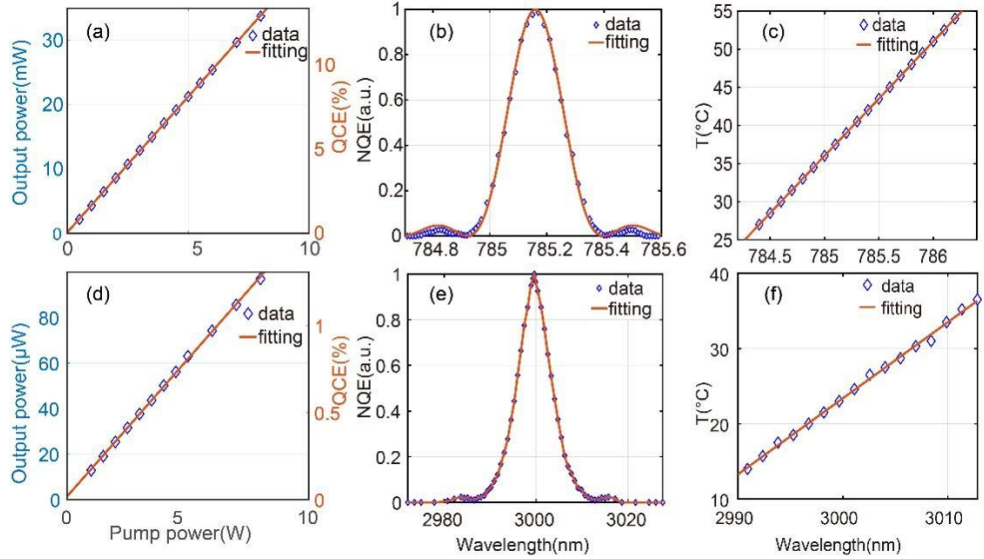

Fig. 3. (a) Relationship between pump power and DFG output power quantum efficiency in PPLN; (b), (c) wavelength and temperature tuning characteristics of DFG in PPLN; (d) Relationship between pump power and SFG output power quantum efficiency in CPPLN; (e), (f) wavelength and temperature tuning characteristics of SFG in CPPLN.

Next, we tested the relevant characteristics of the cascaded process in PPLN. Before the 785T/1064R dichroic mirror, the power of the two pump lasers P1 and P2 was 4.3 W, and the power of the signal beam S was 1 W. The first step of the DFG1 and DFG2 processes generates two light beams with a slight difference in their wavelength of 3000 nm, with each having a power of 17.8 mW. Next, we tested the output spectrum directly after the PPLN crystal and obtained the FWM spectrum in the visible band, as shown in Fig. 4(a). We changed the power of the signal laser S to measure the power of the output sideband light beams SF1 and SF2 in the visible band, where this power is directly proportional to the power of the signal light S, as shown in Fig. 4(b). In addition, the relationship between the wavelength detuning of two pump lasers and the conversion efficiency was tested by varying the wavelength of pump laser P2, as shown in Fig. 4(c). In Fig. 4(c), the red solid line represents the results from a numerical simulation based on the experimental conditions when using Eq. (3) from the theoretical analysis, and the impact of the phase mismatch caused by wavelength detuning is reflected in $h_1(0,\xi_1)$. The measured maximum conversion efficiency of FWM in the visible band was −30.98 dB, and the conversion bandwidth obtained by fitting of the experimental data points was 0.304 nm at 785.16 nm, which is consistent with the results of the theoretical simulation.

Finally, the output spectrum of the mid-infrared FWM was obtained through frequency upconversion, and the results are as shown in Fig. 4(d)–(f). By following the same procedure that was used for the visible-band FWM and varying the pump power of the P1 laser, the conversion efficiency of the mid-infrared FWM versus the pump power P1 was measured, with results as shown in Fig. 4(e); we can see that this is a linear relationship with a slope that is close to 2 in logarithmic coordinates, which is highly consistent with the result from Eq. (5). Moreover, the FWM conversion efficiency versus wavelength detuning was measured and the results are shown in Fig. 4(f), where the red solid line represents the results of the numerical simulation from the theoretical analysis based on Eq. (5). The impact of the phase mismatch caused by wavelength detuning is expressed as $h_i(0,\xi_i)(i=1,2,3)$. The experimental results show that the maximum conversion efficiency for the mid-infrared FWM is −35.45 dB and that the conversion bandwidth is 4.06 nm at 3000 nm; these results are in good agreement with the theoretical predictions and numerical simulations. Two points should be clarified among the experimental measurements here: as the first point, because narrow bandwidth lasers were used in our experiments and the OSA has only limited wavelength resolution, when the wavelength detuning is too small, the powers of the FWM lasers cannot be distinguished from the bottom-level noise of the center light by the OSA; as the second point, when the wavelength detuning is too great, the reduction in conversion efficiency caused by the phase mismatch will also cause the power of the sideband light to be undetectable. Therefore, the wavelength detuning curves have data points missing when the wavelength detuning was either too great or too small, as shown in Fig. 4(c) and 4(f).

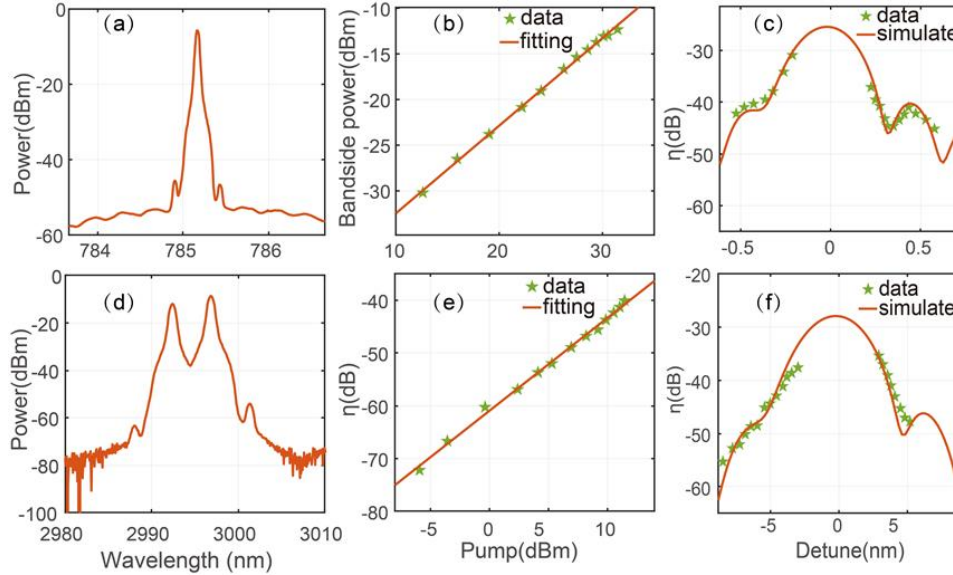

Fig. 4. (a) Output spectrum for the visible FWM; (b) relationships of the signal power $P_{SF1}$ with respect to $P_S$; (c) relationships of the FWM conversion efficiency with respect to the wavelength detuning; (d) output spectrum for the mid-infrared FWM; and (e)-(f) relationships of the FWM conversion efficiency with respect to the pump power and the wavelength detuning, respectively.

In conclusion, we have demonstrated the DPC-DFG+SFG scheme, which generates synthetic FWM in the visible and mid-infrared bands simultaneously via the cascading quadratic nonlinear process in a PPLN crystal. The FWM process in the visible band has a maximum conversion efficiency of as high as −30.98 dB, and the conversion bandwidth obtained at 785.16 nm was 0.304 nm. The mid-infrared FWM parameters were measured using a frequency upconversion step; the maximum conversion efficiency of the mid-infrared FWM process was −35.45 dB, and the conversion bandwidth was 4.06 nm at 3 μm. However, the estimated conversion efficiency of the FWM when generated directly based on the third-order susceptibilities in the PPLN crystal is −143 dB [38-42], which has a conversion efficiency that is lower by 110 dB (at 3 μm) than our experimental result of −30.98 dB in the PPLN crystals. Please refer to the Supplementary Information for further details of this difference. Under the experimental conditions used in this work, because the noise floor of the 1064 nm pump laser when amplified by a fiber amplifier is much greater than the conversion efficiency of −80 dB in the near-infrared band, the FWM spectrum at 1064 nm cannot be obtained and is thus not studied here. As long as the pump power and the crystal conversion bandwidth increase, higher order sidebands are effectively generated in further cascaded processes. This cascading property will be very useful for continuous extension of the frequency ranges of existing lasers, can be used as a powerful tool for generation of free-tuning multiband frequency combs with multi-octave frequency spanning properties, and will lead to significant applications in fields including gas sensing, lidar, and high-precision spectroscopy in the mid-infrared band.

## Author Declarations

**Conflict of Interest:** The authors have no conflicts to disclose.

**Author Contributions**

Li Chen: Conceptualization (equal); Data curation(equal); Formal analysis(equal); Investigation (equal); Methodology (equal); Software (equal); Validation (equal); Writing – original draft (equal); Writing – review & editing (equal). Zheng Ge: Investigation (equal); Methodology (equal); Software (equal). Su-Jian Niu: Formal analysis (equal); Software (equal); Visualization (equal); Writing – original draft (equal). Yin-Hai Li: Data curation (equal); Methodology (equal); Software (equal). Zhao-Qi-Zhi Han: Formal analysis (equal); Software (equal). Yue-Wei Song: Software (equal). Wu-Zhen Li: Investigation (equal); Software (equal). Ren-Hui Chen: Data curation(equal).Ming-Yuan Gao: Visualization (equal). Meng-Yu Xie: Supervision (equal); Visualization (equal). Zhi-Yuan Zhou: Conceptualization (equal); Data curation(equal); Funding acquisition (equal); Methodology (equal); Project administration (equal); Resources (equal); Supervision (equal); Writing – original draft (equal); Writing –review & editing (equal). Bao-Sen Shi: Conceptualization (equal); Funding acquisition (equal); Investigation (equal); Project administration (equal); Resources (equal); Supervision (equal); Writing –review & editing (equal).

**Supplemental document.** See Supplement 1 for supporting content.

**Acknowledgments.** We would like to acknowledge the support from the National Key Research and Development Program of China (2022YFB3607700, 2022YFB3903102), the National Natural Science Foundation of China (NSFC) (11934013, 92065101, 62005068), the Innovation Program for Quantum Science and Technology (2021ZD0301100), and the Space Debris Research Project of China (No. KJSP2020020202). We thank David MacDonald, MSc, from Liwen Bianji (Edanz) (www.liwenbianji.cn/) for editing the English text of a draft of this manuscript.

**Data Availability:** The data that support the findings of this study are available from the corresponding author upon reasonable request.